\documentstyle[12pt]{article}
\newcommand{\be}{\begin{equation}}   \newcommand{\ee}{\end{equation}}
\newcommand{\bd}{\begin{displaymath}} \newcommand{\ed}{\end{displaymath}}
\newcommand{\baa}{\begin{array}{lll}} \newcommand{\eaa}{\end{array}}
\newcommand{\ba}{\begin{eqnarray}}    \newcommand{\ea}{\end{eqnarray}}
\newcommand{\la}{\label}               
\newcommand{\Ds}{\displaystyle}

\def\eg{\hbox{\it e.g.}{}}
\def\MSbar{\relax\ifmmode\overline{\rm MS}\else{$\overline{\rm MS}${ }}\fi}
\def\np#1#2#3{{\em Nucl. Phys.} {\bf B#1} (19#3) #2}

\def\pl#1#2#3{{\em Phys. Lett.} {\bf B#1} (19#3) #2}

\def\pr#1#2#3{{\em Phys. Rev.} {\bf D#1} (19#3) #2}
\def\pr#1#2#3{{\em Phys. Rev.} {\bf D#1} (19#3) #2}

\def\gbar{\relax\ifmmode{\bar{g}}\else{$\bar{g}${ }}\fi}
\def\albar{\relax\ifmmode{\bar{\alpha}}\else{$\bar{\alpha}${ }}\fi}
\def\albars{\relax\ifmmode{\bar{\alpha_s}}\else{$\bar{\alpha_s}${ }}\fi}

\def\GeV{{\rm GeV}}
        
       \def\mgl{m_{\gl}}

\def \as{\relax\ifmmode\alpha_s\else{$\alpha_s${ }}\fi}
\def\gl{\tilde{g}}
\def\asmz{\relax\ifmmode\bar \alpha_s(M_Z)\else{$\bar \alpha_s(M_Z)${ }}\fi}
\def \asQ{\relax\ifmmode\bar \alpha_s(Q)\else{$\bar \alpha_s(Q)${ }}\fi}
\def \asm{\relax\ifmmode\bar \alpha_s(\mu)\else{$\bar \alpha_s(\mu)${ }}\fi}
\def \bal{\relax\ifmmode\bar \alpha \else{$\bar \alpha${ }}\fi}

\begin{document}
{\flushright
     BI - TP 93/75}
{\flushright
     E2-93-336,
     JINR, Dubna}

 \vspace*{1cm}
\begin{center}
{\Large \bf{MASS DEPENDENT $\alpha_S$  EVOLUTION \\
and \\
 \vspace*{0.4cm}
 THE LIGHT GLUINO EXISTENCE}} \\
 \vspace*{1cm}
{\bf D.V. Shirkov \footnote{e-mail: shirkovd@th-head.jinr.dubna.su }\\
 \vspace*{0.5cm}
 {\em Bogoliubov Theoretical Lab., JINR, Dubna, Russia} \\
 \vspace*{0.5cm}
 S.V. Mikhailov \footnote{e-mail: mikhs@theor.jinrc.dubna.su } \\
 \vspace*{0.5cm}
 {\em Bogoliubov Theoretical Lab., JINR, Dubna, Russia} \\
{\em\ and} \\ {\em Sektion Physik, Bielefeld Uni., Bielefeld, Germany}}
\date{}
\end{center}
\thispagestyle{empty}
\setcounter{page}{0}
\begin{abstract}
There is an intriguing discrepancy between \asmz values measured
directly at the CERN $Z_0$-factory and low-energy (at few GeV)
measurements transformed to $Q=M_{Z_0}$ by a massless QCD \asQ
evolution relation.  There exists an attempt to reconcile this
discrepancy by introducing a light gluino $\gl$ in the MSSM. \par
     We study in detail the influence of heavy thresholds on \asQ
evolution.
      First, we consruct the "exact" explicit solution
to the mass-dependent two-loop RG equation for the running \asQ.
This solution describes heavy thresholds smoothly. Second, we
use this solution to recalculate anew \asmz values corresponding to
"low-energy" input data. \par
	Our analysis demonstrates that using  {\it
mass-dependent RG procedure} generally produces corrections of
two types:  Asymptotic correction due to effective shift of
threshold position; 
Local threshold correction only for the case when input experiment
 lies in the close vicinity of heavy particle threshold:
$Q_{expt} \simeq M_h $ \par
   Both effects result in the effective shift of the \asmz values
of the order of $10^{-3}$. However, the second one could be
enhanced when the gluino mass is close to a heavy quark mass. For
such a case the sum effect  could be important for the discussion
of the light gluino existence as it further changes the $\gl$ mass.
\end{abstract}
\newpage
 \section{\bf{Introduction}}

\indent {\it Discrepancy for \asmz value.}
    Recent experiments at CERN and DESY provide the possibility
of a rather accurate (up to few per cent) comparison of data with
the QCD predictions for strong running coupling \asQ evolution
\cite{Bet-92}. The  detailed theoretical analysis results
\cite{ENR-93} in the conclusion that there exists a slight
discrepancy between two groups of experiments, on the one hand,
and ``standard (i.e., treated as a part of SM) QCD", on the other.
\par	So we are confronted with
a new ``Bermuda triangle". This time, unlike the first mismatch
\cite{ugo} connected with GUT speculation at distances of order
$10^{-30}$cm, the new puzzle (if it really exists) concerns physics
at the current frontier of momentum transfer.  \par
	The nature of this new discrepancy admits speculations on
the existence of superpartners with masses lighter than $M_{Z_0}$
\cite{AEN-91},\cite{JK-93},\cite{ENR-93},\cite{RS-93} that is on a
drastic decrease of the SUSY  scale . \par
	Light superpartners, generally, reveal themselves in two ways:
they slow down  \cite{AEN-91}, \cite{JK-93} the running of \asQ  and,
besides, they modify the analysis of experimental data (see, e.g.,
Refs.\cite{ENR-93}, \cite{Heb-93} and \cite{Kolia-93}). \par
\vspace{1mm}
{\it Our program.} The aim of this paper is two-fold.           \par
      The first is to demonstrate the advantage of a mass-dependent
renormalization group (RG) technique for analysing coupling constant
evolution. This technique is based upon the original Bogoliubov
formulation of the RG, the formulation that is not tightly related
to short-distance behaviour, massless approximation and UV
divergencies. It uses as input {\it mass-dependent perturbative
calculation} and takes automatically into account threshold effects.
As this approach usually corresponds to MOM-schemes it can be simply
connected with experimental conditions. It is appropriate to modern
\as measurements at few \GeV \ in decays and DIS where heavy particle
thresholds play a noticeable role. For our discussion we use a
mass-dependent -- "massive" for short -- generalization (see
Eq.(\ref{DV2}) below) of the well-known two-loop UV massless RG
solution for the running coupling
\ba
\Ds \albar^{(2)}(Q) = {\alpha \over 1+\alpha \beta_1\ell +
  \alpha \left(\beta_2/\beta_1\right)\ln(1+\alpha \beta_1\ell)
  }~;\ \ \ \ell= \ln(\frac{Q^2}{\mu^2}) \la{UV}~.
\ea
\par
 	Our second goal consists of discussing of the above-mentioned
discrepancy between the values of  $\asmz$ on the basis of
mass-dependent \asQ evolution. \par
\vspace{1mm}

{\it Results.} To this end we consider details of
this evolution in the threshold vicinity and discover effect
coming from two sources:  \par
	a. Asymptotic correction due to the effective threshold shift
,{\it e.g.}, $Q_{thr}=2m \to 2.30m$. \par
	b. Local threshold correction only for the case when input
experiment momentum lies in the close vicinity of heavy quark (or
gluino) threshold: $q_{expt} \simeq M_h $. \par
     Each effect results in effective shift of the $\asmz$ values of
order of $0.001$. However, the second correction could be enhanced
when gluino mass happens to be close to the heavy quark one\footnote{
One could say that in this case gluino plays a role of a "magnifying
glass" for threshold effect.}. Here the summary effect could be of
order 0.005 that is important for the discussion of a light gluino
hypothesis.

\section{\bf  Mass-dependent \asQ evolution}
 \setcounter{equation}{0}       

\par {\it { Analytic mass-dependent $\asQ $ evolution  on the
two-loop level.}} An exact solution to the one-loop massive RG
equation for the running coupling  $\asQ$ is known from the
mid-fifties \cite{Book},\cite{BSh}. If one starts with the
one-loop mass-dependent perturbative input, then the radiative
correction, after applying RG machinery, goes into
the denominator in the same manner as for the massless case:
\ba \la{RG1}
\Ds \bar \alpha^{(1)}_{pert}(Q^2, m^2, \mu^2) = \alpha -\alpha^2 \cdot
U_1(Q^2, m^2,\mu^2) \ \ \ \to \ \ \  \bar{\alpha}^{(1)}_{RG}(Q) =
{\alpha \over 1+\alpha U_1}~ .
\la{exact1}
\ea
\par On the two-loop  level starting with
\ba
\Ds \albar^{(2)}_{pert}(Q,...)= \alpha -\alpha^2 \cdot U_1+
\alpha^3 \cdot \left( U_1^2 - U_{2}\right) + ...  \nonumber
\ea
one can also obtain \cite{Sh-81} an analytic solution
\ba
\albar^{(2)}_{RG}(Q,m) = \frac \alpha {1 +\alpha U_1 +
\alpha\left(U_2/U_1\right)\ln(1+ \alpha U_1) }.
\la{DV2}
\ea
Here the $l$-loop contribution $U_l$ in a MOM-scheme has a
simple functional structure  \cite{Sh-81}
\ba                           \la{U}
\Ds U_l(Q^2, \ldots ) = I_l(Q^2/m^2)-I_l(\mu^2/m^2)
\ea
and $\mu^2$ is the Euclidean subtraction point.

The RG solution (\ref{DV2}) possesses several remarkable properties.
First, it depends only on the loop-expansion coefficients $U_1$ and
$U_2$ just in the form they appear in the perturbative input.
Second, in the pure massless case with $U_l(Q^2,...) = \beta_l
\ln(Q^2/\mu^2)$  Eq.(\ref{DV2})   precisely corresponds to the
well-known expression (\ref{UV}). Third, being used in QCD, it
smoothly interpolates across the heavy quark thresholds between
corresponding massless expressions with different quark numbers $n_f$.
Finally, expression (\ref{DV2}) can be considered as self-consistent
on the two-loop level, as the error of approximation involved is
of the next-to-next-to-leading, i.e., the three-loop, order (as it
was shown explicitly in Ref.\cite{Sh-92}).   In this sense we shall
refer to it as to an {\it exact two-loop massive solution}.\par
\vspace{2mm}
	{\it One-loop qualitative discussion.}
Result of $U_1$ perturbative QCD calculation  depends on the
renormalization scheme, but {\it not} on the gauge when incoming
fermions are massless (for the general case see \cite{Tar89}). \par
The one-loop massive coefficient $U_1$ ( and the corresponding
$\beta$-function ) was calculated first in \cite{DRG} (see also
\cite{GePo}) on the basis of the ghost-gluon vertex ($U_1^{ggA}$),
then a different expression ($U^{(AAA)}_1$) was obtained
\cite{Nach78} with the triple-gluon vertex. The \asQ evolution based
on the quark-gluon vertex with a massive incoming quarks was
considered in \cite{Tar89}).
The need of a concrete \asQ is determined by the tree structure
of diagrams for a given process. For example, for non-singlet quark
distribution in DIS processes the renormalized triple-gluon vertex
does not occur in the leading $log$ diagrams; nor in quarkonium decays.
So, we shall use only $\asQ^{ggA}$ in the following. \par
	In the QCD case functions $U_l$ consist of a sum of
massless and massive terms
\ba   \la{U1}
U_l(Q^2, \mu^2,.. m^2_h,..)=\beta_l \cdot \ln(\frac{Q^2}{\mu^2}) -
\sum_h \Delta \beta_l^h \cdot H^h_l(Q^2, \mu^2,m^2_h), \\ H^h_l(Q^2,
\mu^2,m^2_h)=I_l^h(\frac{Q^2}{m^2_h})-I_l^h(\frac{\mu^2}{m^2_h}) \nonumber
\ea
with $ 12\pi \beta_1 =11C_A - 2 n_f , ~~
48\pi^2 \beta_2 = 34C_A^2 -38 n_f $ , $n_f$ being the
number of light quarks and the sum extended over all kinds $h$ of
heavy particles. In particular, on the one-loop level one has for
quark contribution $C_A=3,\ \ \Delta \beta_{1}^q =1/6\pi$ and
\ba \la{exactH1}
\Ds I_1^q(z)=2 \sqrt{1+{4\over z}}\left(1-{2\over z}\right)\ln\left(
\sqrt{1+{z\over 4}} + \sqrt{{z\over 4}}\right)+{4\over z} - {5\over 3}
\ \ ; \\
\la{asym1}
 I_1^q(z\to \infty) \to \ln z - {5\over 3}+ O(1/z)\ \ ;\ \ \
I_1^q(z\to 0) \to {z\over 5}\ \ .
\ea
The constant $C_1 = 5/3$ in the asymptotic form
(\ref{asym1}) can be considered as {\it effectively shifting the
threshold position} in the asymptotic logarithmic contribution
\be \la{shift}
I_1^q(Q^2/m^2\to \infty) \to \ln(Q^2/M^2)~~;\ \ \  m \to M= m\exp(5/6)
= 2.301m.
\ee
	One can arrive at an exact solution for $\albar = \albar_s $
 by substituting expressions (\ref{exactH1}) and (\ref{U1})
into Eq.(\ref{exact1}). In the MSSM gluino case
one must appropriately change the $C_A$ factor (due to the Majorana
spinor nature) in the coefficient $\Delta \beta_{1}^q$  in (\ref{U1}).

     Note also that the heavy particles' contribution, as it stands
under the sum on the r.h.s. of Eq.(\ref{U1}), does not satisfy a
decoupling condition at small $Q^2$. It rather corresponds to the
case with $\mu \simeq m_h=m_c , m_b, m_{\gl} << M_Z $. This perfectly
fits our physical situation with $\as (\mu \simeq \mbox{few \GeV) }$
used as input and \asmz being the output. \par
	However, in some cases, (e.g., in discussion of the GUT
consistency problem based upon all three SM couplings $\alpha_i(M_Z)$
experimental values) one should satisfy the decoupling theorem with
respect to all possible particles heavier than given scale ($M_{Z_0}$)
and omit the $I_l^h(\mu^2/m^2_h)$ terms in the sum.
\vspace{1mm}

{\it Simple logarithmic  approximations.} As the exact analytic
form of the fermion one-loop contribution Eq.(\ref{exactH1})
is rather
cumbersome, there was proposed (see \cite{DRG}) a simple
approximation
for $I_1^q$
\be
 \ln(1+\frac{z}{5})=\ln \left(1+\frac{Q^2}{5m^2}\right)  \la{DRG}
\ee               \nonumber
which has reasonable limits at small and large $Q^2$ (compare with
Eq.(\ref{asym1})) and for intermediate $Q^2$ is "accurate within a
few per cent". We see now that a simple modification of
this expression :
\be  
\tilde{I}_1^q(z) \approx   \ln(1+z/5.3)  \la{modDRG}
\ee
restores the exact asymptotic behaviour of $I_1^q(z)$ and leads to
absolute accuracy of the $10^{-2}$ order (for $z \geq 1$)
that is quite enough for our discussion.
\vspace{1mm}

	{\it Two-loop massive contribution to \asQ.}  To avoid a
complicated issue of the gauge dependence, we shall use (see, e.g.,
Refs. \cite{PT-80}, \cite{ShT88}) the transverse (i.e., Landau) gauge.
For our discussion we need an {\it explicit  two-loop mass-dependent
contribution} to the QCD running coupling. This important calculation
for fermion contribution has been performed by Yoshino and Hagiwara
\cite{Jap84}. Unhappily, these authors have published their original
result neither for $U_2$ nor for  contribution to the
$\beta$-function.  The only published expression is the rational
parameterization for the
next-to-leading order  coefficient of the $\beta$-function
$\beta_{ggA}^{(2)}$ in the MOM scheme at the symmetrical Euclidean
point for the ghost-gluon vertex in the Landau gauge:
\ba
(4\pi)^2\beta_{H}^{(2)}\approx  - {38\over 3}\sum_{h}
B_2(\frac{Q^2}{m_h^2})~; \ \ \ B_2(z) =
{0.26995 z^2 - 0.45577 z \over 0.26995 z^2 + 2.1742 z+1}~ .
\la{b2}  \nonumber
\ea
We restore the corresponding approximate expression for $I_2^q(z)$
by elementary integration
: \ba  \la{I2}
 I_2^h(z) &=& 
A\ln(1+z/z_1)+(1-A)\ln(1+z/z_2)~; \\
A&=&1.310 ~ ;\ \ z_1 = 7.579~;\ \ z_2 = 0.4897~. \nonumber
\ea
It is easily seen now, that the "effect of the threshold shift"
increases at the two-loop level
\bd
 I_2^q(z \to \infty) =  \ln z - C_2~;\ \ \ C_2= \left[A\ln(z_1)+
(1-A)\ln(z_2)\right]  \approx  2.872 .
\ed
as compared with the value $C_1 = 5/3$ in (\ref{asym1}). \par
       Using now expression (\ref{I2}) in (\ref{U1}) with
$(4\pi)^2\Delta \beta_{2}^q =38/3$  (for the gluino contribution
$(4\pi)^2\Delta \beta_{2}^{\gl}=48$ \cite{AEN-91},\cite{JK-93}) and
then in Eq.(\ref{DV2}) we obtain an explicit mass-dependent two-loop
evolution law for the QCD running coupling $\albars^{(MOM)}$.

\section{\bf{Analysis of threshold effects in \asQ evolution}}
 \setcounter{equation}{0}
 \par	 {\it Thresholds in massless MS schemes.} The usual way to
account for heavy threshold effects in massless
\asQ  evolution consists in imposing the continuity relations
\ba
    \as[n-1,M_n] = \as[n,M_n]
\la{spline}
\ea
with $M_n$ being the point of "switching on" the n-th quark, i.e.,
 some effective threshold value\footnote{It should be noted, that
this recipe hasn't rigorous substantiation in the field theory and
is devoted to imitate some final results of the mass-dependent
calculation.}. This procedure leads to a "spline-type"
approximation for massless $\albars (Q)$.

      Quite often for the matching point  one takes (see, e.g.,
widely cited Marciano paper\cite{marc}) the n-th particle mass
value $M_n=m_n$. "To be more accurate" sometimes (see,\eg ,
\cite{ENR-93}) one uses the threshold position $M_n=2m_n$.
However, as it follows from Eqs.(\ref{exact1}), (\ref{shift}), the
optimum value of the adjusting parameter
in the relation $M_n=k\cdot m_n$ for fermion polarization
loop (that corresponds to $\beta_{ggA}^{(1)}$) is $k_{opt}^{(1)}=
2.301$\footnote{It is rather curious that in  Ref.
\cite{gott2} this value has been found empirically!} and on the
two-loop level even to $k^{(2)} \approx 4.404$.
    The maximum deviation of the spline approximation
 $\theta(Q - M_h)\ln(Q^2/M^2_h)$ for  $I^h_1$  from the exact
continuous mass-dependent expression (\ref{exactH1}) lies in the
threshold vicinity, being of order of unity.
\par
      However, in real QCD we face with a more complicated situation:
thresholds of heavy quarks and, possibly, of a light gluino are
located rather close to each other. This leads to their mutual
influence that could amplify massive correction near the
threshold region.
\vspace{1mm}

{\it Different schemes at one loop level.} For qualitative
illustrative discussion consider first the one-loop case with a
heavy mass $m_n$. Here it is convenient to represent the QCD running
coupling taken in some i-th subtraction scheme in the form: \par
	a) in the massless case
\be   \label{ai-0}
\frac{1}{\bal_i(Q,m=0)} = \frac{1}{\bal_{i,0}(Q)} = \frac{1}
{\alpha^i_{\mu }}+\beta_{n^*}\ell
+c_{i, n^*} ~; \ \ \  \ell= \ln(\frac{Q^2}{\mu^2}) ~
\ee
with
$$   n^*=n-1  \ \ \mbox{at} \ \ Q \leq m_n ~;~ = n \ \
\mbox{at} \ \ Q \geq m_n~; \ \ \mbox{and}\ \ \beta_n
=\beta_{n-1} - \Delta \beta^n_1 ;  $$
 \par	b) in the massive  case
\be  \label{ai-m}
\frac{1}{\bal_i(Q,m)} = \frac{1}{\alpha^i_{\mu }}+
\beta_{n-1} \ell -\Delta \beta^n_1 \cdot \left[
I^n_1\left(\frac{Q^2}{m^2_n} \right)+C_i \left(\frac{\mu}{m_n} \right)
 \right]
\ee
\noindent where are $c_i, C_i$ some scheme dependent constants. \par
	In particular, for the widely accepted \MSbar scheme we
have two versions: \\
The massless
\be   \label{ams-0}
\frac{1}{\bal_{\MSbar ,o}(Q)} = \frac{1}{\alpha_{\mu ,n^*}}+
\beta_{n^*}\left(\ell-\frac{5}{3}\right)
\ee
and the massive one
\be \label{ams-m}
\frac{1}{\bal_{\MSbar}(Q,m)} = \frac{1}{\alpha_{\mu ,n}}+
\beta_{n-1}\left(\ell-\frac{5}{3}\right) - \Delta \beta^n_1
\cdot \left\{I^n_1\left(\frac{Q^2}{m_n^2} \right)-\ln
\frac{\mu^2}{m^2_n} \right\} ~,
\ee
which evidently relate to each other in the limit $m_n \to 0$
with appropriate change
\be \la{ams0-trans}
\frac{1}{\alpha_{\mu,n-1}} = \frac{1}{\alpha_{\mu,n}}
 - \Delta \beta^n_1 \left[\ln \frac{Q^2}{\mu^2}-\frac{5}{3}\right] ~.
\nonumber  \ee

	Note here that for practical application sometimes one
naively uses the massless Eq.(\ref{ams-0}) in the form
\be   \label{amspop}
\frac{1}{\bal_{\MSbar}(Q)} = \frac{1}{\alpha_{\MSbar , n^*}(\mu)}+
\beta_{n^*}\ell 
\ee
with continuity or {\it matching} condition at some
$Q = M_n \simeq m_n$ that relates coupling constants
\be \la{ams0-match}
\frac{1}{\alpha_{\mu,n-1}} = \frac{1}{\alpha_{\mu,n}}
 - \Delta \beta^n_1 \left[\ln \frac{M_n^2}{\mu^2}-\frac{5}{3}\right] ~.
\ee
	For further discussion we have to consider the \asQ evolution
from some low-energy input scale $\mu=q_{in}\simeq m_{c,b,...}$ up to
$Q=M_{Z_0}$ value. Usually to this end people use massless
Eq.(\ref{ams-0}) which we write down in the spline form
\be   \label{ams-ph}
\frac{1}{\bal_{\MSbar}(Q)}=\frac{1}{\alpha_{\MSbar}(\mu)}+\beta_{n-1}
\ell - \Delta \beta^n_1 \cdot \theta(Q-M_n)\ln \frac{Q^2}{M_n^2}~
\ee
that corresponds to Eq.(\ref{amspop}). \par
	We shall also use the exact smooth massive expression
(\ref{ai-m}) in the MOM scheme
\be   \label{amom-ph}
\frac{1}{\bal_{MOM}(Q,m)} = \frac{1}{\alpha_{MOM}(\mu)}
+\beta_{n-1}\ell - \Delta \beta^n_1 \left\{ I^n_1\left(\frac{Q^2}
{m^2_n}\right) - I_1\left(\frac{\mu^2}{m^2_n} \right)\right\} ~.
\ee

\vspace{2mm}

{\it Qualitative comparison of evolution in different schemes.} The
important point one has to fix before the comparison of the exact smooth
Eq.(\ref{amom-ph}) and massless spline-type Eq.(\ref{ams-ph})
evolution laws is the relation to experimentally measured
quantities. \par
	Being a bit simple-minded  
we suppose that the experimental input
value $\alpha_s(q_{in})$ can be equally treated as
$\alpha_{\MSbar}(q_{in})$ in Eq.(\ref{ams-ph}) or as
$\alpha_{MOM}(\mu = q_{in})$ in Eq.(\ref{amom-ph}). Under this
convention we have
\be   \label{amz-dif}
\frac{1}{\alpha_{MOM}(M_Z)} - \frac{1}{\alpha_{\MSbar}(M_Z)} =
\sum_h \Delta \beta^h_1 \cdot \left\{I^h_1\left(\frac{q_{in}^2}{m_h^2}
\right) + \frac{5}{3} -\ln \frac{M_h^2}{m_h^2} \right\} ~.
\ee
As it follows from this result generally there are two corrections
to the $(\asmz )^{-1}$ value. The first, "asymptotic", one contains
the difference
$\frac{5}{3}-\ln k^2 $ with $1 \leq k=M_h/m_h \leq k^{(1)}_{opt}$
and, at least partially, sometimes is taken into account by the
change of the matching point. However, the second, really "threshold
correction", in the current literature is not taken into account at
all. This contribution
\be   \label{thr-cor}
\Delta_{thr}\frac{1}{\asmz} = \sum_h \Delta \beta^h_1 \cdot
 I^h_1\left(\frac{q_{in}^2}{m_h^2}\right)
\ee
is important for the case when \par {\it input momentum transfer
is close to some heavy particle threshold}. \par
	Postponing the numerical analysis for the last
Section, we make here few comments: \par
	1. As $I_1(z \simeq 1) \simeq 1$ the threshold correction
Eq.(\ref{thr-cor}) for $m_c < q_{in} < m_b$  could be of the order of
$0.05 - 0.10$ which results in the relative effect
$\left(\Delta_{thr}\asmz \right) / \asmz \simeq 10^{-2}$.\par
	2. The abovementioned asymptotic effect takes maximum value
for $k=1$ and for each of the  $c$ and $b$ thresholds produces the
same effect $\Delta_{as}\asmz / \asmz \equiv (1 \div 2)\cdot 10^{-2}$.\par
	3. Second-loop  effects also could be quantitatively estimated
on the base of explicit expression Eq.(\ref{I2}). This leads to the
effective enhancement of one-loop effects by 20 - 30 per cent. \par
\vspace{2mm}

{\it $\MSbar \to MOM \to \MSbar$ transitions.} The other way to take
the threshold into account is to treat the experimental input value
$\alpha_s(q_{in})$ as $\alpha_{\mu,n-1}$ in Eq.(\ref{ams-0}) or
(\ref{ams-m}). The procedure one should use for this situation
contains three steps: \par
{\it(i)} transformation of $\asQ$ from \MSbar to a MOM-scheme at
the low energy scale $q_{in}$;
\par {\it(ii)} evolution of $\alpha_s^{MOM}(q_{in})$ to
$\alpha_s^{MOM}(M_Z)$ by Eq.(\ref{DV2})  that describes smoothly
the gluino  and  heavy quarks thresholds ; \par
{\it(iii)} return transformation from $\alpha_s^{MOM}(M_{Z})$ to
$\alpha_s^{\MSbar}(M_{Z})$ \par
\vspace{1mm}

that is:
\ba               \la{three-step}
& \MSbar \to MOM~~ \mbox{at}~~ q_{in} \simeq m_{c,b}~;& \nonumber \\
&~~~\as^{MOM}(q_{in}) \to ~\as^{MOM}(M_Z)~~\mbox {via Eq.(\ref{DV2})};
& \\ &~~~~~~ MOM \to \MSbar ~\mbox{at}~ Q = M_Z. & \nonumber
\ea
	To connect our equations with data for $\mu=q_{in}$ and $Q=M_Z$
(which are extracted from experiment by using \MSbar --scheme)
 we need relations between
\MSbar (see (\ref{ams-0}) or (\ref{ams-m})) and our version
of MOM  - scheme. \par
	Here is the relation
\ba
{ 1\over \alpha_{MOM}} = { 1\over \alpha_{\MSbar}} - A_1 -
\alpha_{\MSbar}\cdot A_2 ;\ \ \ \ { 1\over \alpha_{\MSbar}} =
{ 1\over \alpha_{MOM}} + A_1 +  \alpha_{MOM}\cdot A_2
\la{recalc}
\ea
where $A_1$ was represented, \eg \ in \cite{Jap84}, with the
constant $\Ds D_1 = (5/3)\beta_1 - (11/12)C_A $,
\ba
A_l(\mu^2/m_h^2,..) =
{1 \over (4 \pi)^l} \left( \sum_h \Delta \beta_l^h \cdot
\left( I_l^h(\mu^2/m_h^2) -  \ln (\mu^2/m_h^2) \right) +D_l \right),
\la{recalc1}
\ea
and the coefficient $A_2$ may be extracted from (\ref{I2})
with $D_2$ being some  constant  dependent on the MOM-scheme chosen.
When using Eqs. (\ref{I2}) at the high energy (HE) scale $Q = M_Z$ we
have to extend
the sum in (\ref{recalc1}) to the  $c,\ b$- quarks and perhaps
gluino. In the low energy (LE) case, quark contributions appear in
the sum as the corresponding thresholds $M_h$ are passed at the scale
$\mu=q_{in}$.
The final result of the three successive steps (\ref{three-step})
is given by formula (\ref{DV2}) with the new $U_l$ (which differ now
from the expression (\ref{U1})) :
\ba   \la{U2N}
U_l^{{New}}(Q^2,\mu^2,..m_h^2,..)= U_l(Q^2,\mu^2,..m_h^2,..) +
A_l(Q^2/m_h^2,..) - A_l(\mu^2/m_h^2,..)
\ea
where all scheme - dependent constants $D_l$ have been canceled.
Substituting expression (\ref{U1}) and (\ref{recalc1}) into
Eq. (\ref{U2N}) we obtain the more detailed expression, where the
gluino contribution is emphasized
\ba   \la{U3N}
U_l^{New}(M_Z^2, q_{in}^2,.. m^2_h,...)=\beta_l \cdot \ln
\frac{M_Z^2}{q_{in}^2} -\Delta \beta_l^{\gl} \cdot \left( \ln {M_Z^2
\over \mgl^2} - I \left({q_{in}^2 \over \mgl^2}\right) \right)\nonumber \\
-  \sum_h \Delta \beta_l^h \cdot \left(\ln(\frac{M_Z^2}{m^2_h}) -
\left[\theta(M_h - q_{in}) I_l^h\left(\frac{q_{in}^2}{m^2_h} \right) +
  \theta(q_{in} - M_h) \ln \frac{q_{in}^2}{m^2_h} \right] \right) .
\ea
Here
\begin{itemize}
\item at the high energy scale all $I_l^h(M_Z^2/m^2_h)$ are replaced
by \MSbar's $\ln(M_Z^2/m^2_h)$ {\bf including} the gluino contribution.
This step corresponds to the scheme
of  $\asmz$ transformation used by the authors of \cite{ENR-93}.
\item at the low energy scale all $I_l^h(q_{in}^2/m^2_h)$ are turned
into  $\ln(q_{in}^2/m^2_h)$
in crossing the quark thresholds in \MSbar - scheme, {\bf excepting}
the gluino contribution $I_l^{\gl}$. The latter is because the
extracted $ \as(q_{in})$ did not include the gluino contribution at
low energy.
\end{itemize}
The results of the calculations of $\asmz$ are represented in the
last column ( MS-MOM-MS) of the Table .

For instructive purposes, let us compare the results of the one-loop
evolution of $\asQ$ in the standard spline approach  Eq.(\ref{ams-ph})
and in the three - step procedure (\ref{three-step}). In the case when
$q_{in}$ is located below the lowest threshold,
the difference of these expressions is
\ba \la{three}
{ 1\over \alpha(M_Z^2)^{New} } - { 1\over \alpha_{\MSbar}(M_Z^2) } =
\sum_h \Delta \beta_1^h \left[
I^h_1 \left(\frac{q_{in}^2}{m^2_h}\right) - \ln \frac{M_h^2}{m_h^2}\right],
\ea
where the RG solution $\asQ^{New}$ correspond to $U_l^{New}$.
In the case when $M_h \gg q_{in}$ and all the particle masses
$m_i$ and $q_{in}$ are separated in logarithmic
scale, one can evidently obtain $M_h \approx m_h,\ \
k \approx 1$ (with some power corrections) for the spline formula
(\ref{ams-ph}). The same conclusion is evidently valid on the two-loop
level.  Comparing the expressions (\ref{three}) and (\ref{amz-dif}) for
the corrections, one can see that there difference includes
the constant $C_1$ only. It is the consequence of the cancelation of
the scheme constants in the procedure (\ref{three-step}).

In the real situation  at low energies when \ \ $m_1 \sim m_2 \ ... \sim
q_{in}$, {\it no universal $M_n$ exist in the spline formula}. Moreover, in
this case any MS scheme calculations become uncertain owing to the
 uncertainty in  fixing the value of $n_f$.

 \section{\bf{LEP data and the light gluino window}}
 \setcounter{equation}{0}
As is known, the discrepancy between low energy $\as$ values and the
recent LEP data has a chance to be resolved by including the light MSSM
gluino $\gl$ \cite{AEN-91},\cite{JK-93},\cite{ENR-93}. \par
	Our main comment is that in a quantitative discussion
of the light gluino mass one should take into account mass effects in
the $\asQ$ evolution in the ways we propose here. As it follows from
our numerical analysis (see three last columns in the Table ) the net
result for the "gluino
existence case" with $m_{\gl} \simeq m_{c,b}$ effectively shifts
\asmz values corresponding to $m_c < q_{in} < m_b$ by few thousandth
which could be physically important.  \par
	At the same time we feel it to be premature to discuss in
detail the light gluino mass owing to rather big experimental errors
of \asmz "direct" measurements. Nevertheless, consider the final
numerical results of the evolution of  $\asQ$ by the different methods,
which are represented in the Table.

In the first column of the Table we have reproduced the \asmz values
following the Bethke recipe (see \cite{Bet-92}), i.e. massless
evolution with the conjunction at $M_n = m_n$.
Note, our first value $(0.116)$ slightly differs from the original one,
and for the eighth value we have used the new experimental data \cite{Topaz}.
The data on $\asmz$,  the gluino effects included are represented in
the next three columns of the Table.

 The results of the
spline MS evolution of $\asQ$, following the massless recipe used by
Ellis et al.
\cite{ENR-93}, are presented in the second column.
In the next two columns we give a new
summary of ``measured" values of $\asmz$ recalculated to scale $M_Z$
using two different smooth mass-dependent
 procedures including the light $\gl$: the first is
based on the pure MOM scheme, the second -- on a three--step procedure
(\ref{three-step}). The application of each of them is dictated by
the way of extracting the $\asQ$ value in low--energy experiments.
One can easily see that the results of both procedures satisfy the
following inequality:
\bd
 \asQ^{New}~~~ \geq ~~~ \asQ^{\MSbar} ~~~ \geq ~~~ \asQ^{MOM}
\ed
An essential theoretical observation is that the net result of
the three--step procedure (\ref{three-step})
is essentially different from the straightforward one
\be \la{Ellis}
\as^{\MSbar}(q_{in}) \to \as^{\MSbar}(M_Z)~~~ \mbox {via spline formulae}
\ee
as  used in \cite{ENR-93} (compare the second and fourth columns in the
 Table).
 The difference from the  results of evolution
in the spline approximation ( with $k=2$, as in \cite{ENR-93})
 being of the order  $5$ per cent is important for the discussion of
the light gluino hypothesis as it happily works in proper direction
raising the ``low-energy \asmz value" (the last column in the Table).

  It should be noted, that the growth of
the cross--section at the Z--peak (see the tenth line in the Table), obtained
 in
\cite{ENR-93}, is due to an extra contribution from gluino part
proportional to $\Delta \beta^{\gl}_1 $ multiplied by $5/3$. At the
same time,  these authors have used the value $k=2$ (instead of
 $exp(5/6)$ !)
for the threshold--shifting parameter.

The additional growth of $\asmz$ in the three--step procedure
 is due to the asymmetry in
treating the gluino contribution  mentioned above. This growth is of
the order  $\Ds \as^2 \Delta \beta^{\gl}_1 \cdot {5/3}
 \approx 0.005$, as it can be seen from the second term in Eq.(\ref{U3N}).

Note at last that the value of $\mgl$ (in the sense of the best $\chi^2$
over all data in the Table) reduces to a value less than $0.5~
\GeV$ in the first naive procedure and grows up to $10~ \GeV$ in the correct
second procedure.

 \section{{\bf Conclusion}}

      1. We believe that our analysis is interesting from the
theoretical point of view, as it demonstrates the possibility of
 rather a simple realization of two-loop {\it mass-dependent} RG
calculation of running \asQ evolution. \par
        Together with the result of recent publication \cite{yuri93}
this opens the door for a systematic consideration of threshold effects
in DIS processes. \par
       Our approach could be of importance for
the discussion of the ``Amaldi et al discrepancy" in GUT scenario.
The point is that the massless \MSbar scheme widely used in this
case ``produces" a systematic shift of the \asQ evolution curve with
respect to a smooth curve of the mass--dependent case. In the GUT
 case this shift is not compensated by the "reverse" scheme transition
$ MOM \to \MSbar ~\mbox{at}~ M_{X,Y} $ at the end.

	2. Our result seems to be interesting
physically as it could contribute into the light gluino discussion.
 Here, the most annoying data are provided
by the first three low energy experimental values in the Table. They
are lying systematically below the others data. On the other hand,
using these data, one can see the intensification of the
``three -- step effect'' due to the light gluino. To this end, one
would compare the columns with and without gluino in the data, we
demonstrate below:
\vspace{1cm}
\begin{center}
\begin{tabular}{|c|c|c|c|c|c|}
                           \hline
 &    Bethke-92    &MS-MOM-MS  & Spline \MSbar &MS-MOM-MS \\
N& \multicolumn{2}{|c|}{}      &\multicolumn{2}{|c|}{}    \\
 &\multicolumn{2}{|c|}{without gluino}&\multicolumn{2}{|c|}{Gluino
  $\mgl=5\GeV$} \\ \hline

 &                 &           &            &\\
$1$&$0.116 \pm 0.005$ &$0.1189$& $0.1309$     &$0.1398$\\
   &               &           &             &\\
$2$&$0.111 \pm 0.006$& $0.1120$ & $0.1233$    & $0.1289$\\
   &               &           &             &\\
$3$&$0.113 \pm 0.005$&$0.1135$ & $0.1253$    &  $0.1304$\\
   &               &           &              &\\  \hline
\end{tabular}
\end{center}

\vspace{1cm}
 However, generally, one should bear in mind that here the
launching platform includes experimental data with remarkable small
error estimates (that seems to be respected by physical community).
The ``SuSy race", we are trying to participate in, is
extremely fashionable at the moment but nobody is sure about the
very existence of a prize. In other words, if further experimental
progress will add arguments in favor of the discrepancy, one would be
open to discussing other physical mechanisms \cite{NOV}.

\newpage
\noindent
\hspace{-5mm}\parbox{\textwidth}{
\begin{center}
 {\bf\Huge Table }  \\
\vspace{10mm}
{\bf Values of $\asmz$ in the standard QCD, and with MSSM gluino $\gl$}
\end{center}
\begin{tabular}{|c|c|c|c|c|c|c|c|}
                           \hline
 &        &        &  Bethke-92    & Spline \MSbar & MOM     &MS-MOM-MS \\
N&Process &$q_{in}$&               & \multicolumn{3}{|c|}{}    \\
 &        & \GeV   & \cite{Bet-92} & \multicolumn{3}{|c|}{Gluino
                $\mgl=5\GeV$} \\ \hline
   \multicolumn{7}{|c|}{} \\ \hline
 &         &  &      &         &            &\\
$1$&$R_{\tau}$\
 [world]\    &$1.78$&$0.116 \pm 0.005$  & $0.131$   &   $0.1274$   &$0.1398$\\
   &         &       &         &            &             &\\
$2$&DIS
 $[\nu]$\     & $5$&$0.111 \pm 0.006$  & $0.123$   &  $0.1204$    &$0.1289$\\
   &       &   &    &         &             &\\
$3$&DIS $[\mu]$
        & $7.1$       &$0.113 \pm 0.005$  & $0.125$  &   $0.1220$ & $0.1304$\\
   &     &     &    &       &              &\\
$4$&$J/\Psi + \Upsilon $
 decays  &$10$      &$0.113 \pm 0.007$  & $0.126$   &$0.1220$     & $0.1291$\\
   &     &          &                    &          &              &\\ \hline
   &       &     &    &       &             &\\
$5$&$p\bar p \to b \bar bX$
          &$20$       & $0.109 \pm 0.014$ &$0.116 $ &  $0.1153$   &$0.1210$ \\
   &       &     &    &       &             &\\
$6$&$e^+ e^- \to [\sigma_{had}]$
        &$34$        & $0.131 \pm 0.012$ &$0.138$  & $0.1376$      &$0.1455$ \\
   &    &        &   &        &             &\\
$7$&$e^+ e^- \to $\ [ev.shapes]
        &$35$    & $0.119 \pm 0.014$ &$0.125$& $0.1247$           &$0.1310$ \\
   &       &     &    &       &             &\\
$8$&$e^+ e^- \to $\ [ev.shapes] (\cite{Topaz})
           &$58$
                 & $0.126 \pm 0.007$ &$0.127$&  $0.1262$    & $0.1325$ \\
   &       &      &     &       &             &\\
$9$&$p\bar p \to W$\ jets
           &$80.6$&$0.121 \pm 0.024$&$0.1215$      & $0.1210$   &$0.1271$ \\
   &       &      &   &       &             &\\ \hline
   \multicolumn{7}{|c|}{} \\ \hline
$10$&$\Gamma(Z^0 \to h)$
      &$91.2$& $0.130 \pm 0.012$&$0.132$ (\cite{ENR-93}) &$0.132$&$0.132$\\
   &       &      &   &       &            &         \\
$11$& $Z^0$\ [ev. shape]
      &$91.2$& $0.120 \pm 0.006$ &$0.124$ (\cite{ENR-93})&$0.124$ &$0.124$\\

   &       &      &        &  or    &   or        & or\\
   &without $\pi^2$ expon. (\cite{ENR-93})
           &      &  &$0.132$ & $0.132$       &$0.132$  \\  \hline
   \multicolumn{3}{|c|}{average of $\asmz$}
                          &0.119   &  &             & $0.131$\\ \hline
\end{tabular}
}
\newpage

{\it Acknowledgements}
The authors are grateful to Prof. Ugo Amaldi exiting our interest
to this problem, and to Drs. L. Avdeev, D. Broadhurst, A. Kataev,
A. Sidorov and O. Tarasov for fruitful discussions of main results.
We are indebted to the Volkswagen foundation and Heisenberg-Landau
Program for financial support as well as to Prof. Jochem Fleischer and
the Physics Department of Bielefeld University for warm hospitality.

\end{document}